\documentclass[12pt]{article}

\usepackage[pdfpagelabels]{hyperref}

\usepackage[usenames,dvipsnames]{color}
\usepackage{graphicx,epsfig,xcolor}

\hypersetup{colorlinks=true, linkcolor=violet, urlcolor=blue, citecolor=blue}

\usepackage{cite}
\usepackage[english]{babel}
\usepackage{amssymb,amsfonts,amsmath}
\usepackage{verbatim}
\usepackage{bbm,bbold,bm}
\usepackage{tensor}
\usepackage{slashed}
\usepackage{braket}

\def\q{\theta}
\def\a{\alpha}
\def\b{\beta}

\def\m{\mu}
\def\n{\nu}
\def\r{\rho}

\def\be{\begin{equation}}
\def\ee{\end{equation}}
\def\ba{\begin{eqnarray}}
\def\ea{\end{eqnarray}}
\def\nb{\nonumber}
\def\p{\partial}

\def\a{\alpha}
\def\b{\beta}

\def\d{\delta}

\def\l{\lambda}
\def\L{\Lambda}
\def\m{\mu}
\def\n{\nu}

\def\r{\rho}

\def\t{\tau}

\def\q{\quad}

\def\mc{\mathcal}

\newcommand{\pr}[1]{\left(#1\right)}
\newcommand{\pq}[1]{\left[#1\right]}
\newcommand{\pg}[1]{\left\{#1\right\}}

\usepackage{chngcntr}
\counterwithin*{equation}{section}

\usepackage{pifont}

\allowdisplaybreaks

\title{\bf On the stability of fracton gravity}

\author{\small Evangelos Afxonidis\footnote{afxonidisevangelos@uniovi.es},
Alessio Caddeo\footnote{caddeoalessio@uniovi.es}, 
Carlos Hoyos\footnote{hoyoscarlos@uniovi.es}, 
Daniele Musso\footnote{mussodaniele@uniovi.es} 
}

\date{}

\begin{document}

\maketitle
\vspace{-22pt}
\begin{center}\it{\small 
Department of Physics and Instituto de Ciencias y Tecnolog\'{\i}as Espaciales de Asturias (ICTEA) Universidad de Oviedo, c/ Leopoldo Calvo Sotelo 18, ES-33007 Oviedo, Spain
\\
}
\end{center}
\vspace{15pt}
\begin{abstract}
We study the stability of fracton gravity, a variant of linearized gravity where the gauge symmetry is restricted to longitudinal diffeomorphisms. These transformations can be connected to a spacetime generalization of dipole symmetry, hence the tag fracton. We find that fracton gravity features an instability in the spin-1 sector corresponding to solutions with growing amplitude. This dynamical instability can be removed by tuning the couplings of the theory. Nonetheless, the Hamiltonian for the spin-1 modes remains always unbounded from below when evaluated on the classical solutions. We find no other sources of instability in the spin-2 or spin-0 sectors. We analyze in detail the canonical formulation  and the constraints arguing that neither auxiliary fields nor gauge-fixing conditions can be employed to remove the problematic vector modes or stabilize them.
\end{abstract}%

\newpage
\tableofcontents

\section{Introduction}
\label{sec:intro}

A possible procedure to recover Einstein's theory of general relativity is to consider a free massless spin-2 field and then introduce interactions that preserve gauge invariance. This can only be pursued consistently if the original spin-2 gauge symmetry is promoted to full diffeomorphism invariance  \cite{Deser:1969wk,Feynman:1996kb,Ortín_2004}, leading as well to universal coupling to matter \cite{Weinberg:1964ew,Weinberg:1980kq}.

One may wonder whether more general theories of massless spin-2 fields could be found by reducing the gauge symmetry. A well-studied case is unimodular gravity, where the symmetry is reduced to transverse diffeomorphisms. The main difference between unimodular and linearized gravity concerns the treatment of the cosmological constant, yet --as long as one assumes a conserved energy-momentum tensor-- the two theories share the physical behavior  \cite{Carballo-Rubio:2022ofy}. A more recent proposal is provided by fracton gravity, where the symmetry is reduced to longitudinal (linearized) diffeomorphisms \cite{Blasi:2022mbl,Bertolini:2023juh,Bertolini:2023sqa,Afxonidis:2023pdq,Rovere:2024nwc}. This was motivated as a relativistic generalization of tensor gauge fields coupled to dipole-preserving theories, first discussed in \cite{Pretko:2016lgv}. One of the most interesting aspects of fracton gravity is that the spin-2 field emerges from a spacetime generalization of \cite{Caddeo:2022ibe} where the dipole gauge symmetry is realized in an internal space. Such spin-2 field can couple to a tensor current which may differ from the energy-momentum tensor, thus allowing for possible non-universality in the coupling to matter \cite{Afxonidis:2023pdq}.

Despite having an interesting structure, it has been argued that massless spin-2 theories with a reduced gauge symmetry are generically unstable, an exception being unimodular gravity (see for instance \cite{Alvarez:2006uu}). Here below we show that this is indeed the case for fracton gravity. For a generic choice of couplings, there is a dynamical instability in the spin-1 sector corresponding to the presence of solutions that grow linearly with time. This instability can be cured by tuning the value of the couplings, but the Hamiltonian remains unbounded from below. As a consequence of unboundedness, the introduction of interactions with matter as well as self-interactions would generically result in a runaway production of the modes that correspond to arbitrarily low energy. We leave as an open question whether interactions can stabilize the theory about a non-trivial vacuum.

The paper is organized as follows. Section \ref{sec:inst} describes the dynamical instability arising due to the mixing among the two spin-1 modes. Section \ref{sec:fixes}, instead, discusses possible attempts at resolving the dynamical instability showing that it can be avoided by means of a suitable choice of the coupling constants. Section \ref{sec:ham} presents the analysis of the canonical structure of the theory, delving into the constraints, the gauge-fixing conditions and the resulting Dirac brackets. Section \ref{sec:spin1en} shows that, despite the removal of the dynamical instability discussed in Section \ref{sec:inst}, the theory suffers from a Hamiltonian which is unavoidably unbounded from below on the classical solutions. Section \ref{sec:conc} concludes the paper and comments on possible stabilization arising from interaction and self-interaction terms. The computations are detailed in the appendices.

\section{Instability of fracton gravity}
\label{sec:inst}

Free fracton gravity theory consists of a symmetric tensor field $h_{\mu\nu}$ with Lagrangian density
\begin{equation}
\label{eq:fracgravLagrangian}
     {\cal L} = (g_1-g_2) h_{\mu\nu} G^{\mu\nu}+ g_2 H_\mu H^\mu \ ,
\end{equation}
where indices are raised and lowered with the flat mostly plus Minkowski metric and we have defined the invariant tensors
\begin{equation}\label{eq:HG}
\begin{split}
& H^\mu=\partial_\sigma h^{\sigma \mu}-\partial^\mu h\ ,\\
& G^{\mu\nu}=\partial^2 h^{\mu\nu}-\partial^\mu \partial^\nu h -(\partial^\mu H^\nu+\partial^\nu H^\mu)+\eta^{\mu\nu}\partial_\sigma H^\sigma\ .   \end{split}
\end{equation}
When $g_2=0$ one recovers the Fierz-Pauli action for standard linearized gravity. For any values of the couplings $g_1$ and $g_2$, the action is invariant under the gauge transformations
\begin{equation}
    \delta h_{\mu\nu}=\partial_\mu \partial_\nu \lambda\ ,
\end{equation}
which corresponds to longitudinal linearized diffeomorphisms
\begin{equation}
    \delta h_{\mu\nu}=\partial_\mu \xi_\nu+\partial_\nu \xi_\mu
    \ , \qquad 
    \xi_\mu=\frac{1}{2}\partial_\mu \lambda\ .
\end{equation}
The reduced gauge symmetry allows for physical massless spin-1 and spin-0 modes, in addition to the spin-2 modes of linearized gravity \cite{Afxonidis:2023pdq}.

The equations of motion obtained from the action above are
\begin{equation}\label{eq:eoms}
  0= (g_1-g_2) G^{\mu\nu}-\frac{g_2}{2}\left(\partial^\mu H^\nu+\partial^\nu H^\mu-2 \eta^{\mu\nu}\partial_\sigma H^\sigma \right)\ .
 \end{equation}
We can expand in Fourier modes with respect to the spatial momentum
\begin{equation}
    h_{\mu\nu}(t,x)
    =
    \int\frac{d^3 k}{(2\pi)^3}\, \tilde h_{\mu\nu}(t,k) \, e^{ik\cdot x}\ .
\end{equation}
There are two spin-1 fields. They correspond to the cases in which one of the indices of $\tilde h_{\mu\nu}$ is transverse while the other one is either temporal or longitudinal (transverse and longitudinal refer to the spatial momentum). Recalling that $\tilde h_{\mu\nu}$ is symmetric, we get
\begin{equation}
    \tilde h^V_i\equiv \left(\delta_{ij}-\frac{k_i k_j}{k^2} \right)\tilde h_{0j}
    \ ,\qquad
    \tilde h^T_i\equiv \left(\delta_{ij}-\frac{k_i k_j}{k^2} \right)\tilde h_{jk} \frac{ik_k}{|k|}\ .
\end{equation}
We have defined $|k|\equiv\sqrt{k^2}$ and the labels $V$ and $T$ refer to the spatial vector and tensor nature of $\tilde h_{0i}$ and $\tilde h_{ij}$, respectively.
The coupled equations for these two spin-1 modes obtained projecting \eqref{eq:eoms} are
\begin{equation}\label{eq:eqsh}
\begin{split}
    0=\,& g_2 \partial_t^2\tilde h^V_i-2(g_1-g_2)k^2 \tilde h^V_i+(g_2-2g_1) |k| \partial_t \tilde h^T_i
    \ ,\\
    0=\,& (g_1-g_2) \partial_t^2 \tilde h^T_i-\frac{g_2}{2} k^2\tilde h^T_i+\left(g_1-\frac{g_2}{2}\right) |k|\partial_t\tilde h^V_i\ ,
\end{split}
\end{equation}
whose solutions are 
\begin{equation}\label{eq:solshVT}
\begin{split}
    \tilde h^V_i=& e^{\pm i |k| t}\left[ A_V^\pm+\frac{g_2-2g_1}{2g_1-3 g_2}(A_T^\pm\mp i A_V^\pm) |k| t\right]\ , \\ 
    \tilde h^T_i=& e^{\pm i |k| t}\left[A_T^\pm+\frac{g_2-2g_1}{2g_1-3 g_2}(A_V^\pm\pm iA_T^\pm)|k| t\right]\ .
    \end{split}
\end{equation}
For $g_2\neq 2g_1$, the amplitude of the modes grows linearly in time, signalling an instability. Note that $g_2 = 2g_1$ would in fact eliminate the coupling terms in \eqref{eq:eqsh}, leading to two independent oscillators.

\section{Removing the dynamical instability}
\label{sec:fixes}

The problematic modes belong to the spin-1 sector and the underlying reason why there is an instability is that we have two coupled harmonic oscillators of the same frequency (see \emph{e.g.} \cite{Alvarez:2006uu}). As \eqref{eq:solshVT} shows, we can avoid it simply by tuning the value of the couplings to $g_2=2g_1$. In principle, we could also try to remove the instability without fixing a relation between $g_1$ and $g_2$, at the prize of introducing additional fields.

Within this second scenario, a possibility one may think of is to introduce a Lagrange multiplier $\Lambda_\mu$ that projects one of the vector modes out. We can write the following gauge-invariant action
\begin{equation}
     {\cal L} = (g_1-g_2) h_{\mu\nu} G^{\mu\nu}+ g_2 H_\mu H^\mu -\Lambda_\mu H^\mu\ ,
\end{equation}
from which we obtain the equations of motion
\begin{align}
    0&= (g_1-g_2) G^{\mu\nu}-\frac{g_2}{2}\left(\partial^\mu H^\nu+\partial^\nu H^\mu-2 \eta^{\mu\nu}\partial_\sigma H^\sigma \right)+\frac{1}{4} \left(\partial^\mu \Lambda^\nu+\partial^\nu \Lambda^\mu-2 \eta^{\mu\nu}\partial_\sigma \Lambda^\sigma \right)
    \ ,\\ 
    0&= H^\mu.
\end{align}
In the spin-1 sector the equations reduce to ($\tilde{\Lambda}_i^V$ is defined similarly to $\tilde{h}_i^V$) 
 \begin{eqnarray}
\label{eq:eq1} (g_1-g_2)\left(\partial_t^2 \tilde{h}^T_i+k^2 \tilde{h}^T_i\right)&=& -\frac{|k|}{4}\tilde{\Lambda}_i^V\ ,\\
\label{eq:eq2} (g_1-g_2)\left(\partial_t^2 \tilde{h}^V_i+k^2 \tilde{h}^V_i\right)&=& \frac{1}{4}\partial_t\tilde{\Lambda}_i^V\ ,\\
\label{eq:eq3} \partial_t \tilde{h}^V_i &= &|k| \tilde{h}^T_i\ .
 \end{eqnarray}
If we combine the time derivative of \eqref{eq:eq2} with \eqref{eq:eq1} multiplied by $|k|$, such that the left hand side cancels out when \eqref{eq:eq3} is used, we arrive at the equation
\begin{equation}
    \partial_t^2\tilde{\Lambda}_i^V+k^2\tilde{\Lambda}_i^V=0\ .
\end{equation}
Therefore, we still have two independent coupled oscillators $\tilde{h}_i^T$ and $\tilde{\Lambda}_i^V$ of the same frequency and the dynamical instability remains.
 
Although the simplest possibility has not worked, it is actually possible to remove the dynamical instability in a gauge-invariant fashion by promoting the Lagrange multiplier to be an auxiliary field $\Lambda_\mu$ with a quadratic term in the action:
\begin{equation}
\label{eq:lagrwithquadraticmultiplier}
     {\cal L} = (g_1-g_2) h_{\mu\nu} G^{\mu\nu}+ g_2 H_\mu H^\mu -\Lambda_\mu H^\mu+\frac{1}{2g_3}\Lambda_\mu\Lambda^\mu\ .
\end{equation}
The equations of motion for this extended action are
\ba
    0&=& (g_1-g_2) G^{\mu\nu}-\frac{g_2}{2}\left(\partial^\mu H^\nu+\partial^\nu H^\mu-2 \eta^{\mu\nu}\partial_\sigma H^\sigma \right)+ \nb
    \\ 
    &+&\frac{1}{4} \left(\partial^\mu \Lambda^\nu+\partial^\nu \Lambda^\mu-2 \eta^{\mu\nu}\partial_\sigma \Lambda^\sigma \right)
    \ ,\\ 
    \Lambda^\mu&=& g_3 H^\mu \ .
\ea
Solving for $\Lambda_\mu$ one gets
\ba
0 &=& (g_1-g_2)(\partial^2 h^{\mu\nu}-\partial^\mu \partial^\nu h)+\left(\frac{g_2}{2}-g_1+\frac{g_3}{4}\right)\left(\partial^\mu H^\nu+\partial^\nu H^\mu\right)+\nb \\
&+&\left(g_1 -\frac{g_3}{2}\right)\eta^{\mu\nu}\partial_\sigma H^\sigma\ ,\\
{\cal L}&=& (g_1-g_2) h_{\mu\nu} G^{\mu\nu}+ \left(g_2-\frac{g_3}{2}\right) H_\mu H^\mu .
\ea
The problematic terms leading to an instability are those proportional to $\partial^\mu H^\nu+\partial^\nu H^\mu$, which can be removed by setting $g_3=2(2g_1-g_2)$. However, making the following redefinitions
\begin{equation}
\label{eq:shiftofparameters}
    g_1'=g_1-\frac{g_3}{2}\ , 
    \qquad 
    g_2'=g_2-\frac{g_3}{2}\ ,
\end{equation}
the resulting equations of motion and Lagrangian are equivalent to the original ones \eqref{eq:eoms} with the new redefined couplings. Thus, the auxiliary field does not really provide an alternative way to remove the instability and one is forced to tune the couplings to the special value $g_2=2g_1$. As we will show later, this makes the Hamiltonian unbounded from below on the classical solutions.

\section{Hamiltonian and classical equations}
\label{sec:ham}

In this section we proceed to construct the Hamiltonian of fracton gravity, then we identify the constraints and the gauge fixing conditions. The canonical momenta and Hamiltonian are defined as usual
\begin{equation}
\begin{split}
 & L=\int d^3x \,\mc{L}
 \ ,\qquad \q
 \Pi^{\mu\nu}=\frac{\delta L}{\delta \partial_t h_{\mu\nu}}
 \ ,\\
 & H=\int d^3 x\,\mc{H}
 \ , \qquad \q \  \, 
 \mc{H}=\Pi^{\mu\nu}\partial_t h_{\mu\nu}-\mc{L}\ .
\end{split}
\end{equation}
In the definition of the canonical momenta and Hamiltonian there is some ambiguity related to boundary terms in the Lagrangian. We choose them in such a way that $\Pi^{0i}=-g_2 H^i$ is gauge invariant and vanishing in the linearized-gravity limit ($g_2\to 0$). Details can be found in Appendix \ref{app:legendre}. The canonical momenta that we obtain are
\begin{subequations}
\label{eq::defmomenta}
\ba
\label{momentum00}
\Pi^{00}  &=& 0 \ , \\
\Pi^{0i}  &=& g_{2} \pr{\p_{t} h_{0i}- \p_{k} h_{ik} - \p_{i} h_{00}  +  \p_{i} h_{kk}} \ , \\
\Pi^{ij} &=& 2 (g_{1} - g_{2})\left(\p_{t} h_{ij} -2 \p_{(i} h_{j) 0} \right)+ \ 2 (2 g_{1} -  g_{2}) \d_{ij} \p_{k} h_{0k}  
- 2 g_{1} \d_{ij} \p_{t} h_{kk} \ ,
\ea
\end{subequations}
while the Hamiltonian density is
\ba
\label{eq::hamiltonianz1}
\mc{H} &=& \frac{1}{4(g_{1}-g_{2})} \pq{\pr{\Pi^{ij}}^{2} - \frac{g_{1}}{2 g_{1} + g_{2}}\pr{\Pi^{ii}}^{2}} + \frac{1}{g_{2}} \pr{\Pi^{0i}}^{2} - 2 h_{00} \Big[(g_{1}- g_{2}) (\p_{i} ^{2} h_{jj} +  \nb \\
&-&  \p_{i}\p_{j} h_{ij} )
+ \p_{i} \Pi^{0i} \Big] 
- h_{0i}\Big\{ 2\partial_j \Pi^{ij}-\frac{g_2}{2g_1+g_2}\partial_i\Pi^{kk} +  \frac{2g_2(g_1-g_2)}{2g_1+g_2}\partial_i \partial_j h_{0j} \Big\} +  \nb \\
&-& 2\Pi^{0i}\partial_ih_{jj}+2\Pi^{0i}\partial_j h_{ij} + V_{\mc{H}} \ . 
\ea
We have used that $\Pi^{i0}=\Pi^{0i}$ and $h_{i0}=h_{0i}$.  We have also introduced $V_{\mc{H}}$, a term involving fields and derivatives with spatial indices only,
\ba
V_{\mc{H}} &=& (g_{1}-g_{2}) \Big[ (\p_{i} h_{jk})^{2} 
-  (\p_{i} h_{kk})^{2} + 2  \p_{i} h_{kk} \p_{j} h_{ij} -  2 (\p_{i} h_{ij})^{2} \Big]\ .
\ea

\subsection{Constraints and Dirac brackets}
\label{sec:constraintsanddiracbrackets}

The condition \eqref{momentum00} corresponds to a primary constraint, restricting the theory to a subspace embedded in the full phase space. The constraint should not change under time evolution, which implies that its Poisson bracket with the Hamiltonian should vanish when restricted to the constrained subspace. This may lead to additional secondary and higher constraints. By repeatedly taking the Poisson bracket with the Hamiltonian, we obtain the following set
\begin{subequations}
\label{eq:constraintsH}
\ba
\Pi^{00} &\approx & 0 \ , \\
\chi_{0} &\equiv& \pg{H,\Pi^{00}} = - 2 \Big[(g_{1} -g_{2})(  \p_{i} ^{2} h_{jj} -\p_{i}\p_{j} h_{ij})
+ \p_{i} \Pi^{0i} \Big] \approx 0\ , \\
\chi_{1} &\equiv& \pg{H,\chi_{0}} = \p_{i} \p_{j} \Pi^{ij} \approx 0\ .
\ea
\end{subequations}
Where $\approx 0$ means that these quantities vanish on the constrained subspace. The bracket  with the last constraint vanishes identically $\pg{H,\chi_{1}} = 0$, so these form a complete and closed set. We will implement the constraints using Dirac brackets. For a set of constraints $\Phi_a$, $a=1,\dots,N$, we first construct the Dirac matrix
\begin{equation}
    C_{ab}(x,y)=\pg{\Phi_a(x),\Phi_b(y)},
\end{equation}
and then define the Dirac bracket of two quantities $X,Y$ as
\begin{equation}
    \pg{X,Y}_D
    \equiv
    \pg{X,Y}-\int d^4x \int d^4 y \pg{X,\Phi_a(x)}C^{-1}(x,y)_{ab}\pg{\Phi_b(y),Y}\ .
\end{equation}
By construction, the Dirac bracket of any quantity with a constraint vanishes identically $\pg{X,\Phi_a(x)}_D=0$. Thus, the Dirac bracket with the Hamiltonian describes the evolution on the constrained subspace.

Note, however, that the constraints \eqref{eq:constraintsH} are all second class 
\be
\pg{\Pi^{00},\chi_{0}} = \pg{\Pi^{00},\chi_{1}} = \pg{\chi_{0},\chi_{1}} = 0 \ .
\ee
Therefore, the Dirac matrix vanishes and it does not admit an inverse. In order to proceed, we need to introduce additional `gauge-fixing' constraints that have non-zero brackets with the ones above, so that the Dirac matrix is invertible. The gauge-fixing conditions are not completely arbitrary, their Poisson brackets with the Hamiltonian should also vanish on the constrained subspace.%
\footnote{In principle one could also add to the Hamiltonian additional terms proportional to the constraints $\mc{H}'=\mc{H}+\tau_{00} \Pi^{00}+\tau_0 \chi_0+\tau_1 \chi_1$, but the $\tau$ multipliers do not appear in any constraints derived with the new Hamiltonian and can then be set to zero. \label{footnotemultipliers}}

A consistent choice of gauge-fixing conditions is (see Appendix \ref{app:usefulresults} for computational details)
\begin{equation}
    \begin{array}{ccl}
    \label{eq:gaugefix}
         K_{00} &=& h_{00}  \ , \\
         K_0 &=&   g_{1} \Pi_{kk} + 2 g_{2} (g_{1}-g_{2})\partial_i h_{0i}   \ , \\
         K_1 &=& g_1\partial_i^2 h_{kk}-\left(g_1-\frac{g_2}{2}\right) \partial_i \partial_j h_{ij} \ .
    \end{array}
\end{equation}
All together, we have $N=6$ constraints, that we identify as 
\begin{equation}
\label{eq:defconstraints}
    \Phi_1=\Pi^{00}
    ,\quad 
    \Phi_2=K_{00}
    \ ,\quad 
    \Phi_3=\chi_0
    \ ,\quad 
    \Phi_4=K_0
    \ ,\quad 
    \Phi_5=\chi_1
    \ ,\quad 
    \Phi_6=K_1\ .
\end{equation}
Defining $\delta_{xy}\equiv \delta^{(3)}(x-y)$, the non-zero components of the Dirac matrix are 
\begin{eqnarray}
    C_{12}(x,y)=-C_{21}(y,x) &=& -\delta_{xy}\ , \\
    C_{34}(x,y)=-C_{43}(y,x) &=& (2g_{1} + g_{2}) (g_1-  g_2) \partial_i^2 \delta_{xy}\ , \\
    C_{46}(x,y)=-C_{64}(y,x) &=& - \frac{ 4g_{1} + g_{2}}{2} g_{1} \partial_i^2\delta_{xy}  \ , \\
    C_{56}(x,y)=-C_{65}(y,x) &=& - \frac{g_2}{2} (\partial_i^2)^2 \delta_{xy} \ .
\end{eqnarray}

\subsection{Equations of motion}

The equations of motion are the Hamilton equations defined in terms of the Dirac brackets constructed above:
\begin{equation}
    \partial_t h_{\mu\nu}(t,x)=-\pg{H,h_{\mu\nu}(t,x)}_D\ ,
    \qquad \partial_t \Pi^{\mu\nu}(t,x)=-\pg{H,\Pi^{\mu\nu}(t,x)}_D\ .
\end{equation}
It is convenient to work in Fourier space
\begin{equation}
    h_{\mu\nu}(t,x)=\int \frac{d^3k}{(2\pi)^3}\Tilde{h}_{\mu\nu}(t,k)e^{ikx}\ ,\qquad  \Pi^{\mu\nu}(t,x)=\int \frac{d^3k}{(2\pi)^3}\Tilde{\Pi}^{\mu\nu}(t,k)e^{ikx}\ .
\end{equation}
The explicit calculations of the equations of motion are straightforward but tedious, we relegate details to Appendix \ref{app:Heqs}. The final outcome is
\begin{subequations}\label{eq:HamEqs}
\ba
\partial_t\Tilde{h}_{0i}&=&\frac{1}{g_2}\Tilde{\Pi}^{0i}+ik_j\Tilde{h}_{ij}  \nb \\
&+& \Big[ \frac{g_1+g_2}{(2g_1+g_2)(g_1-g_2)} \frac{k_j}{k^2}\Tilde{\Pi}^{0j} -  i \frac{g_1}{2g_{1}+g_{2}} \Tilde{h}_{kk}-i \frac{g_1+g_2}{2g_1+g_2} \frac{k_jk_l}{k^2}\Tilde{h}_{jl}\Big] k_i \ ,\\
\partial_t \Tilde{\Pi}^{0i}&=&  i k_j \Tilde{\Pi}^{ij} \nb \\
&-& \frac{g_{2}}{2(2g_1+g_2)}\Big[ i k_i\Tilde{\Pi}^{kk} + i  \frac{k_i}{k^2}k_jk_l\Tilde{\Pi}^{jl} +4 (g_{1}-g_{2}) k_ik_j\Tilde{h}_{0j} \Big] \ , \q \q \\
\partial_t\Tilde{h}_{ij}&=& \frac{1}{2(g_1-g_2)}\Tilde{\Pi}^{ij}+2ik_{(i}\Tilde{h}_{j)0}\nb \\ 
&+& \frac{1}{2(2g_1+g_2)(g_1-g_2)}\Big[-g_1\left( \delta^{ij}+\frac{k_ik_j}{k^2}\right)\Tilde{\Pi}^{kk} +\frac{1}{g_2}(8g_1^2+g_1g_2-g_2^2)\frac{k_ik_j}{k^4 }k_kk_l\Tilde{\Pi}^{kl} \nb \\
&-& 2i g_{2} (g_{1}-g_{2})\frac{k_ik_j}{k^2}k_l\Tilde{h}_{0l} 
- g_1\frac{\delta^{ij}}{k^2}k_kk_l\Tilde{\Pi}^{kl} - 2i g_2 (g_{1}-g_{2}) k_l\Tilde{h}_{0l}\delta^{ij}\Big] \ , \\
\partial_t \Tilde{\Pi}^{ij}&=& 2i k_{(i}\Tilde{\Pi}^{0j)}-2(g_1-g_2)(k^2\Tilde{h}_{ij} -2k_{k} k_{(i} \Tilde{h}_{j)k}) + \nb \\
&-& \frac{2(g_{1}-g_{2})}{2g_{1} + g_{2}} \bigg\{ \left[g_{1} \delta^{ij}+ \pr{g_1 + g_{2}} \frac{k_ik_j}{k^2}\right] k_kk_l\Tilde{h}_{kl} +  \left[\frac{g_{1} + g_{2}}{g_{1}-g_{2}}\frac{k_i k_j}{k^2}+ \frac{g_1}{g_{1}-g_{2}} \delta_{ij}\right] i k_l\Tilde{\Pi}^{0l}  \nb \\
&-& g_{1} \pr{ \delta^{ij} -   \frac{k_ik_j}{k^2} } k^2\Tilde{h}_{kk} \bigg\} \ .
\ea
\end{subequations}

The equations of motion in the Hamiltonian formalism look rather cumbersome. However, several checks can be performed. First, the equations of motion have to be compatible with the constraints $\Phi_a$ discussed in Section \ref{sec:constraintsanddiracbrackets}.
We have verified that this is indeed the case. Moreover, they have to be compatible with the (more compact) equations of motion (\ref{eq:eoms}) derived in the Lagrangian formalism. We have checked that, plugging the momenta (\ref{eq::defmomenta}) in (\ref{eq:HamEqs}), we in fact recover equations (\ref{eq:eoms}).

\subsection{Spin-2 and spin-0 modes}

The spin-2 modes are the simplest to study, they correspond to the transverse traceless components
\begin{equation}
    \Tilde{h}^{TT}_{kk}=0\ ,\qquad k_k \Tilde{h}^{TT}_{ki}=0\ ,\qquad\qquad    \Tilde{\Pi}^{TT}_{kk}=0\ ,\qquad k_k \Tilde{\Pi}^{TT}_{ki}=0\ .
\end{equation}
Their equations of motion are
\ba
\partial_t\Tilde{h}_{ij}^{TT}&=& \frac{1}{2(g_1-g_2)}\Tilde{\Pi}_{ij}^{TT}\ , \\
\partial_t \Tilde{\Pi}_{ij}^{TT}&=& -2(g_1-g_2) k^2\Tilde{h}_{ij}^{TT} \ .
\ea
While the Hamiltonian is
\begin{equation}\label{eq:Hamsp2}
    H_{\rm spin 2}=\int \frac{d^3 k}{(2\pi)^3} \left[\frac{1}{4(g_1-g_2)}\left|\Tilde{\Pi}_{ij}^{TT}\right|^2+(g_1-g_2)k^2\left|\Tilde{h}_{ij}^{TT} \right|^2 \right]\ .
\end{equation}
The solutions to the equations of motion are
\begin{equation}
\Tilde{h}_{ij}^{TT}=e^{\pm i|k| t}\varepsilon_{ij}\ ,\qquad \Tilde{\Pi}_{ij}^{TT}= \pm 2i(g_1-g_2)e^{\pm i|k| t}|k|\varepsilon_{ij}\ .   
\end{equation}
When evaluated on the Hamiltonian, we get
\begin{equation}
    H_{\rm spin 2}=\int \frac{d^3 k}{(2\pi)^3} \left[2(g_1-g_2)k^2\left|\varepsilon_{ij}\right|^2 \right]\ .
\end{equation}
Therefore, the contributions of the spin-2 modes to the energy are positive as long as $g_1-g_2>0$.

The spin-0 modes are rather more complicated, they include the longitudinal and trace parts. A general decomposition is 
\begin{equation}
\begin{split}
    &\Tilde{h}_{ij}^S=\left( \delta_{ij}-\frac{k_i k_j}{k^2}\right) \Tilde{h}^{ST}+\frac{k_i k_j}{k^2}\Tilde{h}^{SL}\ ,\quad \Tilde{h}_{0i}^S=-\frac{ik_i}{|k|}\Tilde{h}_0^L\ ,\quad \Tilde{h}_{00}\ ,\\
    &\Tilde{\Pi}_{ij}^S=\left( \delta_{ij}-\frac{k_i k_j}{k^2}\right) \Tilde{\Pi}^{ST}+\frac{k_i k_j}{k^2}\Tilde{\Pi}^{SL}\ ,\quad \Tilde{\Pi}_{0i}^S=-\frac{ik_i}{|k|}\Tilde{\Pi}_0^L,\quad \Tilde{\Pi}_{00}\ ,\\
\end{split} 
\end{equation}
The constraints fix
\begin{eqnarray}
    \Tilde{\Pi}_{00} &=& 0, \qquad \Tilde{h}_{00} =0\ ,\\
    \Tilde{\Pi}_0^L &=& 2(g_1-g_2)|k|\Tilde{h}^{ST}\ ,\\
    \Tilde{h}_0^L &=& -\frac{g_1}{g_2(g_1-g_2)|k|}\Tilde{\Pi}^{ST}\ ,\\
    \Tilde{\Pi}^{SL} &=& 0\ ,\\
    \Tilde{h}^{SL} &=& -\frac{4g_1}{g_2}\tilde{h}^{ST} \ .
\end{eqnarray}
With these constraints the spatial components simplify to
\begin{equation}
    \Tilde{\Pi}_{ij}^S=\left( \delta_{ij}-\frac{k_i k_j}{k^2}\right)\Tilde{\Pi}^{ST},\qquad  \Tilde{h}_{ij}^S=\left( \delta_{ij}-\frac{k_i k_j}{k^2}-\frac{4g_1}{g_2}\frac{k_i k_j}{k^2}\right)\Tilde{h}^{ST}.
\end{equation}
It follows that there is a single spin-0 component, with amplitude $\Tilde{h}^S=\tilde{h}^{ST}$, $\tilde{\Pi}^S=\tilde{\Pi}^{ST}$. The Hamilton equations for the spin-0 modes are
\begin{eqnarray}
    \partial_t \Tilde{h}^S &=& \frac{1}{2(g_1-g_2)}\Tilde{\Pi}^S\ ,\\
    \partial_t \Tilde{\Pi}^S &=& -2(g_1-g_2) k^2 \Tilde{h}^S.
\end{eqnarray}
The solutions are
\begin{equation}
    \tilde{h}^S=e^{\pm i |k| t}\varepsilon_0 \ , \q \quad \tilde{\Pi}^S=\pm 2i|k|(g_1-g_2) e^{\pm i |k| t}\varepsilon_0\ .
\end{equation}
Evaluating the Hamiltonian \eqref{eq::hamiltonianz1}, we get
\begin{equation}\label{eq:Hamsp0}
\begin{split}
    H_{\rm spin\; 0}=&\int \frac{d^3 k}{(2\pi)^3}\left[\frac{2g_1+g_2}{2g_2(g_1-g_2)}\left|\Tilde{\Pi}^S\right|^2+\frac{2(g_1-g_2)(2g_1+g_2)}{g_2}k^2\left|\Tilde{h}^S\right|^2\right]\\
    =&\int \frac{d^3 k}{(2\pi)^3}\left[\frac{4(2g_1+g_2)(g_1-g_2)}{g_2}k^2\left|\varepsilon_0\right|^2 \right].
\end{split}
\end{equation}
Therefore, the contributions to the energy from the spin-0 modes is always positive as long as $(g_1-g_2)(2g_1+g_2)/g_2>0$. Together with the condition $g_1-g_2>0$ from the spin-2 modes, this means $(2g_1+g_2)/g_2>0$. Then, for $g_1>0$ both the spin-2 and spin-0 have positive energy if $0<g_2<g_1$ or $g_2<-2g_1$. On the other hand, if $g_1<0$ then the contributions are positive for $g_2<0$ with $|g_2|>|g_1|$.

\section{Instability in Hamiltonian formalism}
\label{sec:spin1en}

Let us first identify the dynamical instability. Projecting the Hamilton equations \eqref{eq:HamEqs} on the spin-1 sector, we get
\begin{subequations}\label{eq:HamEqsVec}
\ba
\partial_t\Tilde{h}_{i}^V&=&\frac{1}{g_2}\Tilde{\Pi}_i^V+|k|\Tilde{h}_{i}^T \ ,\\
\partial_t \Tilde{\Pi}_i^V&=&  |k| \Tilde{\Pi}_i^T  \ , \q \q \\
\partial_t\Tilde{h}_i^T&=& \frac{1}{2(g_1-g_2)}\Tilde{\Pi}_i^T-|k|\Tilde{h}_i^V\ , \\
\partial_t \Tilde{\Pi}_i^T&=& -|k|\Tilde{\Pi}_i^V \ .
\ea
\end{subequations}
Combining them, we obtain the following second order equations
\ba
(\partial_t^2 +k^2)\Tilde{\Pi}_i^{T,V} &= &0 \ , \\ 
\label{eq:hiV}
(\partial_t^2 +k^2)\Tilde{h}_i^{V} &= & |k|\left( \frac{1}{g_2}+\frac{1}{2(g_1-g_2)}\right)\Tilde{\Pi}_i^T \ , \\
\label{eq:hiT}
(\partial_t^2 +k^2)\Tilde{h}_i^{T} &= & -|k|\left( \frac{1}{g_2}+\frac{1}{2(g_1-g_2)}\right)\Tilde{\Pi}_i^V \ .
\ea
The canonical momenta enter as oscillatory sources in the equations for $\tilde h_i^V$ and $\tilde h_i^T$, with the same frequency as that of the homogeneous part. This leads to solutions whose amplitude features the linear in time behavior observed in \eqref{eq:solshVT}. Such circumstance is avoided only when the the coefficients of the momenta in \eqref{eq:hiV} and \eqref{eq:hiT} vanish, that is, for $g_2=2g_1$.

Let us now evaluate the contribution of the spin-1 sector to the Hamiltonian. Note that the spin-1 modes do not enter in the constraints \eqref{eq:constraintsH} and \eqref{eq:gaugefix}, thus we do not need to analyze them. Taking this into account, the Hamiltonian for the spin-1 modes is 
\begin{equation}\label{eq:Hams1}
    H_{\rm spin 1}=\int \frac{d^3k}{(2\pi)^2}\left[\frac{1}{2(g_1-g_2)}\left|\Tilde{\Pi}_i^T\right|^2+\frac{1}{g_2}\left|\Tilde{\Pi}_i^V\right|^2+|k|\left((\Tilde{\Pi}_i^V)^* \Tilde{h}_i^T- (\Tilde{h}_i^V)^* \Tilde{\Pi}_i^T+c.c.\right)\right].
\end{equation}
From equations \eqref{eq:HamEqsVec} with $g_2=2g_1$, we obtain the solutions
\begin{equation}\label{eq:stablesols}
    \Tilde{\Pi}_i^V=e^{\pm i|k| t} \beta_i\ ,\ \ \Tilde{\Pi}_i^T=\pm ie^{\pm i|k| t} \beta_i\ ,\ \ \Tilde{h}_i^V=e^{\pm i |k| t}\alpha_i\ ,\ \ 
    \Tilde{h}_i^T=e^{\pm i |k| t}\left(\pm i \alpha_i-\frac{1}{2 g_1 |k|}\beta_i\right)\ .
\end{equation}
Introducing these into \eqref{eq:Hams1} (with $g_2=2g_1$) we arrive at
\begin{equation}\label{eq:Hams1ab}
\begin{split}
    H_{\rm spin 1}=&\int \frac{d^3k}{(2\pi)^2}\left[-\frac{1}{g_1}|\beta_i|^2\pm 2 i|k|(\beta_i^*\alpha_i-\alpha_i^*\beta_i)\right]\\
    &=\int \frac{d^3k}{(2\pi)^2}\left[-\frac{1}{g_1}\left|\beta_i\mp 2i g_1 |k| \alpha_i\right|^2+4g_1 k^2|\alpha_i|^2\right].
    \end{split}
\end{equation}
From this expression it is manifest that the Hamiltonian has a direction within the space of solutions along which it can take arbitrarily negative values, for either $g_1>0$ or $g_1<0$. As a consequence, coupling the spin-1 sector of the theory to matter would in principle produce a runaway emission of spin-1 modes. On the other hand, the Hamiltonians for the spin-2 \eqref{eq:Hamsp2} and spin-0 \eqref{eq:Hamsp0} modes would be unbounded from below for $g_1>0$ and bounded for $g_1<0$, so additional instabilities can still be evaded in the other sectors.

The Hamiltonian is unbounded from below also when one considers generic values for $g_1$ and $g_2$. Completing the squares in \eqref{eq:Hams1},
\begin{equation}\label{eq:Hams1B}
\begin{split}
    H_{\rm spin 1}=&\int \frac{d^3k}{(2\pi)^2}\left[\frac{1}{2(g_1-g_2)}\left|\Tilde{\Pi}_i^T-2(g_1-g_2)|k| \Tilde{h}_i^V\right|^2+\frac{1}{g_2}\left|\Tilde{\Pi}_i^V+g_2 |k| \Tilde{h}_i^T\right|^2\right.\\
    &\left.-2(g_1-g_2)k^2\left|\Tilde{h}_i^V\right|^2-g_2k^2\left|\Tilde{h}_i^T \right|^2\right].
    \end{split}
\end{equation}
The general solution to the equations of motion is
\begin{equation}
\begin{split}
    &\Tilde{\Pi}_i^V=e^{\pm i|k| t} \beta_i\ ,\ \ \Tilde{\Pi}_i^T=\pm ie^{\pm i|k| t} \beta_i\ ,\ \ \Tilde{h}_i^V=e^{\pm i |k| t}\left(\alpha_i+\frac{2g_1-g_2}{4(g_1-g_2)g_2}\beta_i t \right)\ ,\\ 
    &\Tilde{h}_i^T=e^{\pm i |k| t}\left(\pm i \alpha_i-\frac{2g_1-3g_2}{4(g_1-g_2)g_2|k|}\beta_i\pm i \frac{2g_1-g_2}{4(g_1-g_2)g_2}\beta_i t \right)\ .
\end{split}
\end{equation}
Which reduces to \eqref{eq:stablesols} when $g_2=2g_1$. Evaluating the Hamiltonian \eqref{eq:Hams1B} on this solution, one finds
\begin{equation}
\begin{split}
    H_{\rm spin 1}=&\int \frac{d^3k}{(2\pi)^2}\left[\frac{1}{(g_1-g_2)}\left|\beta_i\right|^2\pm 2i |k|(\beta_i^*\alpha_i-\alpha_i^*\beta_i)\right]\ ,
    \end{split}
\end{equation}
which leads to the same situation as in \eqref{eq:Hams1ab} upon replacing $g_1\to g_2-g_1$.

We have repeated the canonical analysis including the auxiliary field introduced in the Lagrangian in Section~\ref{sec:fixes}. As anticipated, its effect is just a shift of the couplings, thus leading to the same conclusion about the instability. Details can be found in Appendix~\ref{app:lagmult}.

\section{Conclusions}
\label{sec:conc}

We showed that quadratic fracton gravity is unstable. For generic values of the couplings, the theory features solutions that grow linearly in time. When the couplings are tuned to avoid the dynamical mixing responsible for the linearly-growing solutions, the Hamiltonian is still unbounded from below on the classical solutions. Introducing quartic or higher self-interactions could perhaps stabilize the theory about a non-trivial ground state. If so, the stable vacuum would present a non-zero condensate of spin-1 modes, hence breaking Lorentz invariance spontaneously. 

An important part of the analysis above concerns the classical canonical formulation of fracton gravity entailing the identification of the constraints as well as the necessary gauge-fixing conditions. Remarkably, we found only scalar conditions while a vector gauge-fixing was earlier proposed within the Lagrangian formalism \cite{Bertolini:2023juh}. Specifically, such vector gauge fixing condition corresponds to adding to the action a term like
\begin{equation}
    \label{eq:ddgf}
    \mc{L}_{\text{gf}}=-\frac{1}{2\xi}(\partial^\sigma h_{\sigma\mu}+\kappa \partial_\mu h)^2
    \ ,\qquad 
    \kappa\neq -1\ .
\end{equation}
This is similar to the de Donder gauge fixing for linearized gravity and it leads to consistent equations and propagators. The apparent tension with the canonical analysis above is resolved by realizing that \eqref{eq:ddgf} is actually a scalar gauge fixing condition in disguise, in fact it can be rewritten as
\begin{equation}
    \mc{L}_{\text{gf}}
    =
    -\frac{1}{2\xi}(H_\mu+(\kappa+1) \partial_\mu h)^2
    =
    -\frac{1}{2\xi} \left[H_\mu H^\mu+(\kappa^2-1) (\partial_\mu h)^2-2(\kappa+1)h \partial^\sigma\partial^\mu h_{\mu\sigma}\right]\ ,
\end{equation}
where we have used integration by parts in the last term to emphasize that it only involves the spin-0 part. The vector part proportional to $H_\mu H^\mu$ just shifts the values of the couplings $g_1$ and $g_2$, in an analogous way to the auxiliary field in Section \ref{sec:fixes}; in particular, it does not affect the stability analysis above.

\section*{\large Acknowledgements}

We thank Carlo Maria Becchi, Erica Bertolini, Alberto Blasi, Matteo Carrega, Antón Faedo, Silvia Fasce, Carlo Iazeolla and Nicola Maggiore for relevant and enjoyable discussions. The work of A.C. is supported
by Ministerio de Ciencia e Innovaci\'on de Espa\~na under the program Juan de la Cierva-formaci\'on. The work of E.A. is supported by the Severo Ochoa fellowship PA-23-BP22-170. This work is partially supported by the AEI and the MCIU through the
Spanish grant PID2021-123021NB-I00.

\appendix


\section{From Lagrangian to Hamiltonian formalism}
\label{app:legendre}

In this appendix, we start from the fracton gravity Lagrangian (\ref{eq:fracgravLagrangian}) and derive the momenta conjugate to $h_{0i}$ and $h_{ij}$, and the Hamiltonian. Let us rewrite the Lagrangian adding all the boundary terms compatible with $\Pi^{00}=0$ and with a $g_{1}$-independent $\Pi^{0i}$,%
\footnote{
The Lagrangian in \eqref{Lagrangiantimeandspacesplit} can be obtained from 
\ba
    {\cal L} 
    &=& 
    g_1\left(\partial_\alpha h \partial^\alpha h
    - \partial_\alpha h_{\beta\gamma} \partial^\alpha h^{\beta\gamma}
    - 2 \partial_\alpha h \partial_\beta h^{\alpha\beta}
    + 2 \partial_\alpha h^{\alpha\beta} \partial^\gamma h_{\beta\gamma}\right) \nb
    \\  
    &+& g_2 \left(
    \partial_\alpha h_{\beta\gamma} \partial^\alpha h^{\beta\gamma}
    - \partial_\alpha h^{\alpha\beta} \partial^\gamma h_{\beta\gamma}\right)\ ,
\ea
supplemented with the following boundary terms
\ba
\label{calB}
    {\cal B} &=& 
    4(g_1-zg_2)\,\partial_{[i}\left(h_{0i}\partial_{0]}h_{jj}\right)
    +4(g_1+vg_2)\,\partial_{[0}\left(h_{0i}\partial_{j]}h_{ij}\right) \nb
    \\  
    &+&(g_1-g_2) \pq{ \partial_{\alpha}\left(h_\beta^{\ \gamma}\partial_{\gamma}h^{\alpha\beta}\right) - \partial_{\gamma}\left(h_\beta^{\ \gamma}\partial_{\alpha}h^{\alpha\beta}\right) }\ .
\ea
The anti-symmetrizations guarantee that the boundary terms do not yield any term with two derivatives acting on the same field. Since $\partial_0 h_{00}$ appears only in the last term in \eqref{calB}, that one is the boundary term needed to get $\Pi_{00}=0$.
}
\ba
\label{Lagrangiantimeandspacesplit}
\mc{L} &=& (g_{1}- g_{2}) \pr{\p_{0} h_{ij} }^{2}
-g_{1} \pr{\p_{0} h_{kk} }^{2}
+ g_{2} (\p_{0} h_{0i})^{2} + 4 g_{1} \p_{0} h_{ii} \p_{j} h_{0j} + \nb \\
&-& 2 (2g_{1} - g_{2} u) \p_{0} h_{ij} \p_{i} h_{0j} + 2 v g_{2} \p_{0} h_{0j} \p_{i} h_{ij} + 2 g_{1} \p_{i} h_{00} \pr{\p_{j} h_{ij} - \p_{i} h_{jj}}
\nb \\
&-&g_1 \p_{i} h_{0j} \p_{j} h_{0i}
-g_1 \pr{\p_{i} h_{0i}}^2
\nb \\
&+& g_{2} \big[ 2z \pr{\p_{0} h_{0i} \p_{i} h_{jj} - \p_{0} h_{jj} \p_{i} h_{0i}} -2 \p_{0} h_{0i} \p_{i} h_{00} + (\p_{i}h_{00})^{2} 
+ \nb \\
&+& \a \p_{i} h_{0j} \p_{j} h_{0i} + \b (\p_{i} h_{0i})^{2} \big] 
+ 2(g_{1}- g_{2}) \pr{\p_{i} h_{0j}}^{2}
- V_{\mc{L}}  \ .
\ea
The boundary terms are parameterized by $\pg{\a,\b,u,v,z}$ which are real and have to satisfy
\be
\a + \b = 1 \ , \q \q \q u+v=1 \ ,
\ee
whereas $z$ is unconstrained. Note that these parameters affect the definition of the momenta. Moreover, $V_{\mc{L}}$ is a potential term involving only spatial indices
\ba
V_{\mc{L}} &=& (g_{1}-g_{2}) \big[ (\p_{i} h_{jk})^{2} - \p_{i} h_{jk} \p_{j} h_{ik} \big]
- g_{1}  (\p_{i} h_{kk})^{2} + \nb \\
&+& 2 g_{1} \p_{i} h_{kk} \p_{j} h_{ij} - g_{1} (\p_{i} h_{ij})^{2} \ .
\ea
From (\ref{Lagrangiantimeandspacesplit}) we find the momenta
\begin{subequations}
\label{eq::defmomentaApp}
\ba
\Pi^{00}  &=& 0 \ , \\
\label{momentum0i}
\Pi^{0i}  &=& g_{2} \pr{\p_{0} h_{0i} - \p_{i} h_{00} + v \p_{j} h_{ij} + z \p_{i} h_{jj}} \ , \\
\Pi^{ij} &=& 2 (g_{1} - g_{2}) \p_{0} h_{ij} - 2 (2 g_{1} - u g_{2}) \p_{(i} h_{j) 0} + \ 2 (2 g_{1} - z g_{2}) \d_{ij} \p_{k} h_{0k} + \nb \\
&-& 2 g_{1} \d_{ij} \p_{0} h_{kk} \ .
\label{momentumij}
\ea
\end{subequations}
Taking the Legendre transform of the Lagrangian, we obtain the Hamiltonian
\ba
\label{hamiltonianwithfreeparameters}
\mc{H} &=& \frac{1}{4(g_{1}-g_{2})} \pq{\pr{\Pi^{ij}}^{2} - \frac{g_{1}}{2 g_{1} + g_{2}}\pr{\Pi^{ii}}^{2}} + \frac{1}{g_{2}} \pr{\Pi^{0i}}^{2} - 2 h_{00} \Big[(g_{1}-z g_{2}) \p_{i} ^{2} h_{jj} +  \nb \\
&-& (g_{1}+v g_{2}) \p_{i}\p_{j} h_{ij}
+ \p_{i} \Pi^{0i} \Big] 
- \frac{h_{0i}}{g_{1}- g_{2}} \Big\{ (2g_{1} - u g_{2}) \p_{j} \Pi^{ij} + \nb \\
&+& \frac{g_{2} \pq{g_{1}(u-z-2)+g_{2} z}}{(2g_{1}+g_{2})} \p_{i} \Pi^{kk} 
+ \frac{g_{2} (u-2)}{2} \pq{g_{2} (u+2) - 4 g_{1}} \p_{j}^{2} h_{0i}
+ \nb \\
&+& a (g_1-g_2) \p_{i}\p_{j} h_{0j} \Big\} - 2 z \Pi^{0i} \p_{i} h_{jj}- 2 v \Pi^{0i} \p_{j} h_{ij} + V_{\mc{H}} \ .
\ea
We denoted with $a$ the quite cumbersome coefficient
\be
a = \frac{g_2 \left[g_2^2 \left(u^2+2\right)+g_1^2 (4-8 u)+2 g_2 g_1 (2
   u-1)\right]}{2 \left(g_1-g_2\right) \left(2 g_1+g_2\right)}+\frac{2
   g_2 z \left(g_2 u+4 g_1\right)}{2 g_1+g_2}-\frac{3 g_2^2 z^2}{2
   g_1+g_2} \ .
\ee
Besides, we gathered in $V_{\mc{H}}$ the terms involving fields and derivatives with spatial indices only. Notice that $V_{\mc{H}} \neq V_{\mc{L}}$, as we obtain further purely spatial terms from the square of $\Pi^{0i}$ in (\ref{momentum0i}). In fact, 
\ba
V_{\mc{H}} &=& (g_{1}-g_{2})  (\p_{i} h_{jk})^{2} 
- (g_{1} - g_{2} z^{2}) (\p_{i} h_{kk})^{2} + 2 (g_{1} + vz g_{2}) \p_{i} h_{kk} \p_{j} h_{ij} + \nb \\
&-&  \big[ 2 g_{1}- g_{2} (1+ v^{2}) \big] (\p_{i} h_{ij})^{2}\ .
\ea
A choice for the parameters that conveniently simplifies the Hamiltonian and makes $\Pi^{0i}$ invariant under the gauge transformations $\d h_{\m \n} = \p_\m \p_\n \l$ is
\be
\label{eq:parcho}
z = 1 \ , \q \q \q u = 2 \ , \q \q \q v \equiv 1 - u =  -1 \ .
\ee
The parameter $\a = 1 -\b$ does not play a relevant role and we can take $\a=1$ and $\b=0$. Adopting \eqref{eq:parcho}, the Hamiltonian becomes
\ba
\label{eq::hamiltonianz1app}
\mc{H} &=& \frac{1}{4(g_{1}-g_{2})} \pq{\pr{\Pi^{ij}}^{2} - \frac{g_{1}}{2 g_{1} + g_{2}}\pr{\Pi^{ii}}^{2}} + \frac{1}{g_{2}} \pr{\Pi^{0i}}^{2} - 2 h_{00} \Big[(g_{1}- g_{2}) (\p_{i} ^{2} h_{jj} +  \nb \\
&-&  \p_{i}\p_{j} h_{ij} )
+ \p_{i} \Pi^{0i} \Big] 
- h_{0i}\Big\{ 2\partial_j \Pi^{ij}-\frac{g_2}{2g_1+g_2}\partial_i\Pi^{kk} +  \frac{2g_2(g_1-g_2)}{2g_1+g_2}\partial_i \partial_j h_{0j} \Big\} +  \nb \\
&-& 2\Pi^{0i}\partial_ih_{jj}+2\Pi^{0i}\partial_j h_{ij} + V_{\mc{H}} \ , 
\ea
with
\ba
V_{\mc{H}} &=& (g_{1}-g_{2}) \Big[ (\p_{i} h_{jk})^{2} 
-  (\p_{i} h_{kk})^{2} + 2  \p_{i} h_{kk} \p_{j} h_{ij} -  2 (\p_{i} h_{ij})^{2} \Big]\ .
\ea

\section{Derivation of the Hamilton equations}
\label{app:Heqs}

In this appendix we provide the detailed derivation of the equations of motion in the Hamiltonian formalism. For constrained systems, the Hamilton equations are defined by means of the Dirac brackets. We denote with $X_\a$ the fields $\pg{h_{0i}, \Pi_{0i}, h_{ij}, \Pi_{ij},h_{00},\Pi^{00}}$. In this compact notation, the Hamilton equations read
\begin{eqnarray}
    \partial_t X_\alpha(y)=-\{ H,X_\alpha(y)\}_D\ .
\end{eqnarray}
Adopting an analogous compact notation for the constraints $\Phi_a$, the Dirac brackets between two generic dynamical quantities $Y$ and $Z$ are defined through
\begin{equation}
    \pg{Y,Z}_D=\pg{Y,Z}-\int d^4x \int d^4 y \pg{Y,\Phi_a(x)}C^{-1}(x,y)_{ab}\pg{\Phi_b(y),Z} \ ,
\end{equation}
where $C^{-1}_{ab}(x,y)$ is the inverse of the Dirac matrix, defined as
\begin{subequations}
\label{eq:notationforeom}
\begin{equation}
    C_{ab}(x,y)=\pg{\Phi_a(x),\Phi_b(y)} \ .
\end{equation}

We assume translational invariance and express the Hamilton equation in Fourier space. Let us first define
\ba 
\label{eq:defS}
 S_{b\alpha}(z,y) & \equiv & \{ \Phi_b(z), X_\alpha(y)\}=S_{b\alpha}(z-y)\ \ , \\
 \label{eq:defDelta}
 \Delta_a(x) &\equiv &\{H,\Phi_a(x)\}  \ , \\
 \label{eq:defP}
 P_\alpha(y)&\equiv&\{H, X_\alpha(y)\} \ ,
\ea
\end{subequations}
and denote with a tilde their Fourier transform. For instance,
\begin{equation}
     S_{b\alpha}(x-y)=\int \frac{d^3k}{(2\pi)^3} e^{ik\cdot (x-y)} \tilde{S}_{b\alpha}(k) \ .
\end{equation}
The equations of motion in Fourier space then read
\begin{eqnarray}
i \omega X_\a 
=\tilde{P}_\alpha(k)-\tilde{\Delta}_a(k)\tilde{C}_{ab}^{-1}(-k)\tilde{S}_{b\alpha}(-k)\ .
\end{eqnarray}


\subsection{Some useful results}
\label{app:usefulresults}

We here provide explicit expressions for the quantities given in equations (\ref{eq:notationforeom}).
The constraints $\Phi_a$ that we consider here are those of Section \ref{sec:constraintsanddiracbrackets}, namely
\begin{equation}
    \begin{array}{ccl}
    \label{eq::sixcontraintsz1}
         \Phi_1 &=& \Pi^{00}  \ , \\
         \Phi_2 &=& h_{00}  \ , \\
         \Phi_3 &=& \partial_i\Pi^{0i}+(g_1 -g_{2}) \partial_i^2 h_{kk}-(g_1-g_2) \partial_i \partial_j h_{ij} \ ,  \\
         \Phi_4 &=&   g_{1} \Pi_{kk} + 2 g_{2} (g_{1}-g_{2})\partial_i h_{0i}   \ , \\
         \Phi_5 &=& \partial_i\partial_j \Pi^{ij}  \ , \\
         \Phi_6 &=& g_1\partial_i^2 h_{kk}-\left(g_1-\frac{g_2}{2}\right) \partial_i \partial_j h_{ij} \ .
    \end{array}
\end{equation}
The non-vanishing entries of the Dirac matrix read
\begin{subequations}
\begin{eqnarray}
    \tilde{C}_{12}(k)=-\tilde{C}_{21}(-k) &=& -1 \ ,\\
    \tilde{C}_{34}(k)=-\tilde{C}_{43}(-k) &=& - (2g_{1} + g_{2}) (g_1-  g_2) k^2 \ ,\\
    \tilde{C}_{46}(k)=-\tilde{C}_{64}(-k) &=& \frac{ 4g_{1} + g_{2}}{2} g_{1} k^2 \ ,\\
    \tilde{C}_{56}(k)=-\tilde{C}_{65}(-k) &=& - \frac{g_2}{2} k^4 \ ,
\end{eqnarray}
\end{subequations}
while the non-vanishing entries of the Dirac matrix are
\begin{subequations}
\begin{eqnarray}
    \tilde{C}^{-1}_{12}(k)=-\tilde{C}^{-1}_{21}(-k) &=& 1 \ , \\
    \tilde{C}^{-1}_{34}(k)=-\tilde{C}^{-1}_{43}(-k) &=& \frac{1}{(2g_1+g_2)(g_1-g_2)}\frac{1}{k^2} \ , \\
    \tilde{C}^{-1}_{35}(k)=-\tilde{C}^{-1}_{53}(-k) &=&\frac{(4g_1+g_2)}{(2g_1+g_2)(g_1-g_2)}\frac{g_1}{g_2} \frac{1}{k^{4}} \ , \\
    \tilde{C}^{-1}_{56}(k)=-\tilde{C}^{-1}_{65}(-k) &=& \frac{2}{g_2 k^4} \ .
\end{eqnarray}
\end{subequations}
The Poisson brackets of the Hamiltonian with the fields $X_{\a} =\{ h_{0i}, \Pi_{0i}, h_{ij}, \Pi_{ij}, h_{00}, \Pi^{00}\}$ with $\a=1,2,3,4,5,6$ read (see definition \eqref{eq:defP}):
\begin{subequations}
\ba
\Tilde{P}_1 &=& -\left( \frac{1}{g_2}\Tilde{\Pi}_{0i}+ik_j\Tilde{h}_{ij}+i k_i \Tilde{h}_{00} -i k_i\Tilde{h}_{kk}\right)  \ ,\\
\Tilde{P}_2 &=& -\left( ik_j\Tilde{\Pi}_{ij}-\frac{2g_2(g_1-g_2)}{2g_1+g_2}k_i k_j \Tilde{h}_{0j}-\frac{g_2}{2(2g_1+g_2)}ik_i\Tilde{\Pi}^{kk}\right)  \ ,\\
\Tilde{P}_3 &=& -\frac{1}{2(g_1-g_2)}\left( \Tilde{\Pi}_{ij}-\frac{g_1}{2g_1+g_2}\delta_{ij}\Tilde{\Pi}^{kk}\right)-2i k_{(i}\Tilde{h}_{j)0}+i\frac{g_2 }{2g_1+g_2} \delta_{ij} k_k\Tilde{h}_{0k}  \ , \q\\
\Tilde{P}_4 &=&-2i k_{(i}\Tilde{\Pi}_{j)0}+2ik_k\Tilde{\Pi}^{0k}\delta_{ij}+ 2(g_1-g_2)\Big( k^2\Tilde{h}_{ij} - k^2 \Tilde{h}_{kk}\delta_{ij} + k_k k_l \Tilde{h}_{kl}\delta_{ij} + \nb \\
&+& k_ik_j\Tilde{h}_{kk} -2 k_k k_{(i}\Tilde{h}_{j)k} + k^2\Tilde{h}_{00}\delta_{ij} - k_ik_j \Tilde{h}_{00}\Big) \ ,  \\
\Tilde{P}_6 & = &  -2\left(i k_i \tilde{\Pi}^{0i}-(g_1-{g_2}(k^2\Tilde{h}_{kk}-k_ik_j\Tilde{h}_{ij})\right) \ .
\ea
\end{subequations}
The Poisson brackets of the constraints with the fields are (see definition \eqref{eq:defS}):
\begin{subequations}
\ba
\Tilde{S}_{31} &=& -\frac{i}{2}k_i  \ ,\\
\Tilde{S}_{42}&=&ig_2(g_1-g_2)k_i  \ ,\\
\Tilde{S}_{43}&=&-g_1\delta_{ij}  \ ,\\
\Tilde{S}_{53}&=&k_ik_j  \ ,\\
\Tilde{S}_{34}&=&-(g_1-g_2)\left(k^2\delta_{ij}-k_ik_j\right)  \ ,\\
\Tilde{S}_{64}&=&-g_1k^2\delta_{ij}+\left(g_1-\frac{g_2}{2}\right)k_ik_j  \ ,\\
\Tilde{S}_{15} & = &  -1  \ ,\\
\Tilde{S}_{26} & = &  1  \ .
\ea
\end{subequations}
The Poisson brackets of the Hamiltonian with the fields are (see definition \eqref{eq:defDelta}):
\begin{subequations}
\ba
\Tilde{\Delta}_1&=&2(g_1-g_2)\left( k^2\Tilde{h}_{ii}-k_ik_j\Tilde{h}_{ij}\right)-2ik_i\Tilde{\Pi}^{0i} \ ,\\
\Tilde{\Delta}_3&=&\frac{1}{2}k_ik_j\Tilde{\Pi}^{ij}  \ ,\\
\Tilde{\Delta}_4 &=& 2(g_1+g_2)ik_i\Tilde{\Pi}^{0i}-2(g_1^2-g_2^2)\left(k^2\Tilde{h}_{kk}-k_ik_j\Tilde{h}_{ij}\right)+ \\
&+& 2\left(2g_1^2-g_2^2-g_1g_2\right)k^2\Tilde{h}_{00} \ , \\
\Tilde{\Delta}_6&=&-\frac{2g_1-g_2}{4(g_1-g_2)}k_ik_j\Tilde{\Pi}^{ij}+\frac{g_2^2}{2(2g_1+g_2)}ik^2k_j\Tilde{h}_{0j}+\frac{g_1g_2}{4(g_1-g_2)(2g_1+g_2)}k^2\Tilde{\Pi}^{kk}  \ . \q \q \q
\ea
\end{subequations}

The set of constraints (\ref{eq::sixcontraintsz1}) is consistent if their Poisson brackets with the Hamiltonian vanishes. In order to perform this check, the following results are useful.
\begin{subequations}
\label{partialbracketsz1}
\ba
\pg{\p_{i} h_{0i}, H} &=& \frac{1}{g_{2}} \p_{i} \Pi^{0i} + \p_{i} \p_{j} h_{ij} + \p_{i}^{2} h_{00}-\partial_i^2h_{kk} \ ,  \\ 
\pg{\p_{i} \p_{j} h_{ij}, H}&=&  \frac{1}{2(g_{1}-g_{2})} \pq{\p_{i} \p_{j} \Pi^{ij} - \frac{g_{1}}{2 g_{1} + g_{2} } \p_{i}^{2}\Pi^{kk}} + \frac{4g_1+  g_2}{2g_1+g_2} \p_{i}^{2} \p_{j} h_{0j}  \ , \q \\
\pg{\p_{i}^{2} h_{kk}, H} &=&  - \frac{1}{2(2g_{1}+g_{2})} \p_{i}^{2} \Pi^{kk} + \frac{4g_1-g_2}{2g_1+g_2} \p_{i}^{2} \p_{j} h_{0j}  \ , \\
\pg{\Pi^{kk}, H} &=& -4\partial_i \Pi^{0i}-2(g_1-g_2)\left(\partial_i^2h_{kk}-2\partial_i^2h_{00}-\partial_i\partial_jh_{ij}\right)   \ .
\ea
\end{subequations}

\subsection{Equations of motion}

The equations of motion for the fields $X_{\a} = \{h_{0i}, \Pi_{0i}, h_{ij}, \Pi_{ij},h_{00},\Pi^{00}\}$ are given by
\ba
-i\omega \Tilde{h}_{0i}&=&\frac{1}{g_2}\Tilde{\Pi}^{0i}+ \Big[ \frac{g_1+g_2}{(2g_1+g_2)(g_1-g_2)} \frac{k_j}{k^2}\Tilde{\Pi}^{0j} -  i \frac{g_1}{2g_{1}+g_{2}} \Tilde{h}_{kk}-i \frac{g_1+g_2}{2g_1+g_2} \frac{k_jk_l}{k^2}\Tilde{h}_{jl}\Big] k_i + \nb \\
&+&ik_j\Tilde{h}_{ij} \ ,
\ea
\ba
-i \omega \Tilde{\Pi}^{0i}&=& - \frac{g_{2}}{2(2g_1+g_2)}\Big[ i \Tilde{\Pi}^{kk} + i  \frac{k_jk_l}{k^2}\Tilde{\Pi}^{jl} +4 (g_{1}-g_{2}) k_j\Tilde{h}_{0j} \Big] k_i + i k_j \Tilde{\Pi}^{ij} \ , \q \q
\ea
\ba
-i\omega \Tilde{h}_{ij}&=& \frac{1}{2(g_1-g_2)}\Tilde{\Pi}^{ij}+2ik_{(i}\Tilde{h}_{j)0} + \frac{1}{2(2g_1+g_2)(g_1-g_2)}\Big[-g_1\left( \delta^{ij}+\frac{k_ik_j}{k^2}\right)\Tilde{\Pi}^{kk} + \nb \\
&+&\frac{1}{g_2}(8g_1^2+g_1g_2-g_2^2)\frac{k_ik_j}{k^4 }k_kk_l\Tilde{\Pi}^{kl}-2i g_{2} (g_{1}-g_{2})\frac{k_ik_j}{k^2}k_l\Tilde{h}_{0l} + \nb \\
&-&g_1\frac{\delta^{ij}}{k^2}k_kk_l\Tilde{\Pi}^{kl} - 2i g_2 (g_{1}-g_{2}) k_l\Tilde{h}_{0l}\delta^{ij}\Big] \ ,
\ea
\ba
-i\omega \Tilde{\Pi}^{ij}&=& 2i k_{(i}\Tilde{\Pi}^{0j)}-2(g_1-g_2)k^2\Tilde{h}_{ij} +4(g_1-g_2)k_{k} k_{(i} \Tilde{h}_{j)k} + \nb \\
&-& \frac{2(g_{1}-g_{2})}{2g_{1} + g_{2}} \bigg\{ \left[g_{1} \delta^{ij}+ \pr{g_1 + g_{2}} \frac{k_ik_j}{k^2}\right] k_kk_l\Tilde{h}_{kl} +  \left[\frac{g_{1} + g_{2}}{g_{1}-g_{2}}\frac{k_i k_j}{k^2}+ \frac{g_1}{g_{1}-g_{2}} \delta_{ij}\right] i k_l\Tilde{\Pi}^{0l}  \nb \\
&-& g_{1} \pr{ \delta^{ij} -   \frac{k_ik_j}{k^2} } k^2\Tilde{h}_{kk} \bigg\}  \ , \\ 
-i\omega \Tilde{h}_{00} &=& 0 \ , \\
-i\omega \Tilde{\Pi}^{00} &=& 0\ .
\ea

\section{Canonical analysis with the Lagrange multiplier}
\label{app:lagmult}

In this appendix, we perform the canonical analysis for the case with an auxiliary field $\L_\m$ used to enforce the condition $\L_\m = g_3 H_\m$ as discussed in Section \ref{sec:fixes}. In particular, we show the consistency of the Hamiltonian and the constraints with the redefinition of the couplings $g_1,g_2$ as in (\ref{eq:shiftofparameters}).

The Lagrangian that we consider here is (\ref{eq:lagrwithquadraticmultiplier}) improved with the same boundary terms as in Appendix \ref{app:legendre}, that is
\ba
\mc{L} &=& (g_{1}- g_{2}) \pr{\p_{0} h_{ij} }^{2}
-g_{1} \pr{\p_{0} h_{kk} }^{2}
+ g_{2} (\p_{0} h_{0i})^{2} + 4 g_{1} \p_{0} h_{ii} \p_{j} h_{0j} + \nb \\
&-& 4 (g_{1} - g_{2}) \p_{0} h_{ij} \p_{i} h_{0j} - 2  g_{2} \p_{0} h_{0j} \p_{i} h_{ij} + 2 g_{1} \p_{i} h_{00} \pr{\p_{j} h_{ij} - \p_{i} h_{jj}}
\nb \\
&-&g_1 \p_{i} h_{0j} \p_{j} h_{0i} - g_1 \pr{\p_{i} h_{0i}}^2 + g_{2} \big[ 2 \pr{\p_{0} h_{0i} \p_{i} h_{jj} - \p_{0} h_{jj} \p_{i} h_{0i}} -2 \p_{0} h_{0i} \p_{i} h_{00} \nb \\
&+&   (\p_{i}h_{00})^{2} 
+  \p_{i} h_{0j} \p_{j} h_{0i}  \big] 
+ 2(g_{1}- g_{2}) \pr{\p_{i} h_{0j}}^{2} - \L_{0} \pr{\p_{i} h_{0i} - \p_{0} h_{kk}} \nb \\
&+&    \L_{i} \pr{- \p_{0} h_{0i} + \p_{i} h_{00}} - \frac{1}{2 g_{3}} \L_{0}^{2} + \frac{1}{2 g_{3}} \L_{i}^{2} - V_{\mc{L}} \ ,
\ea
where
\ba
V_{\mc{L}} &=& (g_{1}-g_{2}) \big[ (\p_{i} h_{jk})^{2} - \p_{i} h_{jk} \p_{j} h_{ik} \big]
- g_{1}  (\p_{i} h_{kk})^{2} + \nb \\
&+& 2 g_{1} \p_{i} h_{kk} \p_{j} h_{ij} - g_{1} (\p_{i} h_{ij})^{2}  - \L_{i} \pr{\p_{j} h_{ij} - \p_{i} h_{kk}} \ .
\ea
The momenta read
\begin{subequations}
\ba
\Pi^{0} _{\L}  &=& 0 \ , \\
\Pi^{i} _{\L}  &=& 0 \ , \\
\Pi^{00}  &=& 0 \ , \\
\Pi^{0i}  &=& g_{2} \pr{\p_{0} h_{0i} - \p_{i} h_{00} -  \p_{j} h_{ij} +  \p_{i} h_{kk}} - \frac{\L_{i}}{2}  \ , \\
\Pi^{ij} &=& 2 (g_{1} - g_{2}) \p_{0} h_{ij} - 4 ( g_{1} -  g_{2}) \p_{(i} h_{j) 0} + \ 2 (2 g_{1} -  g_{2}) \d_{ij} \p_{k} h_{0k} + \nb \\
&-& 2 g_{1} \d_{ij} \p_{0} h_{kk} +  \d_{ij} \L_{0}  \ .
\ea
\end{subequations}
The Hamiltonian is
\ba
\label{hamiltonianwithmultipliers}
\mc{H} &=& \frac{1}{4(g_{1}-g_{2})} \pq{\pr{\Pi^{ij}}^{2} - \frac{g_{1}}{2 g_{1} + g_{2}}\pr{\Pi^{ii}}^{2}} + \frac{1}{g_{2}} \pr{\Pi^{0i}}^{2} - 2 h_{00} \Big[(g_{1}- g_{2}) (\p_{i} ^{2} h_{kk} +  \nb \\
&-&  \p_{i}\p_{j} h_{ij} )
+ \p_{i} \Pi^{0i} \Big] 
- h_{0i}\Big\{ 2\partial_j \Pi^{ij}-\frac{g_2}{2g_1+g_2}\partial_i\Pi^{kk} +  \frac{2g_2(g_1-g_2)}{2g_1+g_2}\partial_i \partial_j h_{0j} \Big\} +  \nb \\
&-& 2\Pi^{0i}\partial_ih_{kk}+2\Pi^{0i}\partial_j h_{ij} +\frac{1}{2 \left(2 g_1+g_2\right)} \Pi^{kk} \L_{0} 
+ \frac{1}{g_{2}} \Pi^{0i} \L_{i} + \pq{\frac{1}{2 g_3} - \frac{3}{4 \left(2 g_1+g_2\right)} } \L_{0}^{2} +
\nb \\ 
&+& \pq{ \frac{1}{4 g_{2}} - \frac{1}{2 g_3}} \L_{i} ^{2} -\frac{2 \left(g_1-g_2\right) }{2 g_1+g_2} \p_{i} h_{0i} \L_{0} + \t_{i} \Pi^{i} _{\L} + \t_{0} \Pi^{0} _{\L}
+ V_{\mc{H}} \ ,
\ea
where
\ba
V_{\mc{H}} &=& (g_{1}-g_{2}) \Big[ (\p_{i} h_{jk})^{2} 
-  (\p_{i} h_{kk})^{2} + 2  \p_{i} h_{kk} \p_{j} h_{ij} -  2 (\p_{i} h_{ij})^{2} \Big]\ ,
\ea
and where we have included multipliers $\t_{\m}$  for the constraints associated to the momenta of $\L_\m$, since they will show later in the brackets with the Hamiltonian and are necessary to recover the full equations.

The constraints read
\begin{subequations}
\ba
\Pi^{0} _{\L} &=& 0 \ , \\
\Pi^{i}_{\L} &=& 0 \ , \\
\Pi^{00} &=& 0 \ , \\
\r_{0} &\equiv& \pg{\mc{H},\Pi^{0}_{\L}} = \frac{1}{2(2g_{1}+g_{2})} \pq{\Pi^{kk} - 4(g_{1}-g_{2}) \p_{i} h_{0i} - 3  \L_{0}} + \frac{1}{g_3} \L_{0}  \ , \\
\r_{0i} &\equiv& \pg{\mc{H},\Pi^{i}_{\L}} = \frac{1}{g_{2}} \Pi^{0i} + \pq{ \frac{1}{2 g_{2}} - \frac{1}{g_3} }\L_{i}  \ , 
\label{eq:constraintlambdaipi0i}
\\
\chi_{0} &\equiv& \pg{\mc{H},\Pi^{00}} = - 2 \Big[\p_{i} \Pi^{0i} + (g_{1}-g_{2})\pr{  \p_{i} ^{2} h_{kk} -\p_{i}\p_{j} h_{ij} }
 \Big] \ , \\
\r_1 &\equiv&\pg{\mc{H},\r_{0}} = \nb \\
&=& \frac{1}{ g_2 (2g_1 + g_2)} \bigg[ 2 g_1 \p_i \Pi^{0i}  + g_2 (g_1-g_2) \pr{\p_i \p_j h_{ij} - \p_i ^2 h_{kk}} +  (g_1-g_2) \p_i \L_i
\bigg]+ \nb \\
&-& \pq{\frac{1}{g_3} - \frac{3 }{2(2g_{1}+g_{2})}} \t_{0} \ , \\
\r_{1i} &\equiv&\pg{\mc{H},\r_{0i}} = \nb \\
&=&  - \frac{1}{2 g_{2}} \pq{2\partial_j \Pi^{ij}-\frac{g_2}{2g_1+g_2}\partial_i\Pi^{kk} +  \frac{2(g_1-g_2)}{2g_1+g_2}\pr{ 2 g_2 \partial_i \partial_j h_{0j} -  \p_{i} \L_{0} } } + \nb  \\
&-& \pq{\frac{1}{2 g_{2}} -  \frac{1}{g_3}} \t_{i} \ ,\\
\chi_{1} &\equiv& \pg{\mc{H},\chi_{0}} = \p_{i} \p_{j} \Pi^{ij} \ .
\ea
\end{subequations}
The conditions $\rho_1 \approx 0$ and $\rho_{1i} \approx 0$  fix the multipliers $\t_\m$,
\begin{subequations}
    \ba
\t_0 &=& \frac{2 g_3 (g_1-g_2)}{g_2\left[ 2(2g_1+g_2)-3 g_3 \right]}\bigg[ \frac{2 g_1}{g_1-g_2} \p_i \Pi^{0i}  + g_2  \pr{\p_i \p_j h_{ij} - \p_i ^2 h_{kk}} +  \p_i \L_i
\bigg] \ , \\
\t_i &=&  \frac{g_3}{2g_2-g_3}\pq{2\partial_j \Pi^{ij}-\frac{g_2}{2g_1+g_2}\partial_i\Pi^{kk} +  \frac{2(g_1-g_2)}{2g_1+g_2}\pr{ 2 g_2 \partial_i \partial_j h_{0j} -  \p_{i} \L_{0} } } \ . \q \q
    \ea
\end{subequations}
A consistent set of gauge fixing conditions is
\begin{subequations}
\ba
G_1 &=& h_{00}  \ , \\
G_2 &=& \Pi^{kk} + \frac{2 \left(g_1-g_2\right) \left(2 g_2- g_3\right)}{2 g_1- g_3}\partial_i h_{0i} \ ,\\
G_{3} &=& \partial_i^2 h_{kk}- \pr{\frac{1}{2}+\frac{g_1-g_2}{2g_1- g_3}} \partial_i \partial_j h_{ij}\ .
\ea
\end{subequations}

Solving (\ref{eq:constraintlambdaipi0i}) for $\L_i$ and plugging it in the Hamiltonian (\ref{hamiltonianwithmultipliers}), we find the same Hamiltonian (\ref{eq::hamiltonianz1app}) once the shift 
\begin{equation}
    g_1'=g_1-\frac{g_3}{2}\ , 
    \qquad 
    g_2'=g_2-\frac{g_3}{2}\ ,
\end{equation}
is taken into account. It can be easily checked that the same occurs for the constraints and the gauge-fixing conditions, they become \eqref{eq::sixcontraintsz1} with the shift in the couplings.

\bibliographystyle{utphys}
\bibliography{bibliography}

\providecommand{\href}[2]{#2}\begingroup\raggedright\begin{thebibliography}{10}

\bibitem{Deser:1969wk}
S.~Deser, ``{Self-interaction and gauge invariance},''
  \href{http://dx.doi.org/10.1007/BF00759198}{{\em Gen. Rel. Grav.} {\bfseries
  1} (1970) 9--18}, \href{http://arxiv.org/abs/gr-qc/0411023}{{\ttfamily
  arXiv:gr-qc/0411023}}.

\bibitem{Feynman:1996kb}
R.~P. Feynman, {\em {Feynman lectures on gravitation}}.
\newblock 1996.

\bibitem{Ortín_2004}
T.~Ortín, {\em Gravity and Strings}.
\newblock Cambridge Monographs on Mathematical Physics. Cambridge University
  Press, 2004.

\bibitem{Weinberg:1964ew}
S.~Weinberg, ``{Photons and Gravitons in $S$-Matrix Theory: Derivation of
  Charge Conservation and Equality of Gravitational and Inertial Mass},''
  \href{http://dx.doi.org/10.1103/PhysRev.135.B1049}{{\em Phys. Rev.}
  {\bfseries 135} (1964) B1049--B1056}.

\bibitem{Weinberg:1980kq}
S.~Weinberg and E.~Witten, ``{Limits on Massless Particles},''
  \href{http://dx.doi.org/10.1016/0370-2693(80)90212-9}{{\em Phys. Lett. B}
  {\bfseries 96} (1980) 59--62}.

\bibitem{Carballo-Rubio:2022ofy}
R.~Carballo-Rubio, L.~J. Garay, and G.~Garc\'\i{}a-Moreno, ``{Unimodular
  gravity vs general relativity: a status report},''
  \href{http://dx.doi.org/10.1088/1361-6382/aca386}{{\em Class. Quant. Grav.}
  {\bfseries 39} no.~24, (2022) 243001},
  \href{http://arxiv.org/abs/2207.08499}{{\ttfamily arXiv:2207.08499 [gr-qc]}}.

\bibitem{Blasi:2022mbl}
A.~Blasi and N.~Maggiore, ``{The theory of symmetric tensor field: From
  fractons to gravitons and back},''
  \href{http://dx.doi.org/10.1016/j.physletb.2022.137304}{{\em Phys. Lett. B}
  {\bfseries 833} (2022) 137304},
  \href{http://arxiv.org/abs/2207.05956}{{\ttfamily arXiv:2207.05956
  [hep-th]}}.

\bibitem{Bertolini:2023juh}
E.~Bertolini, A.~Blasi, A.~Damonte, and N.~Maggiore, ``{Gauging fractons and
  linearized gravity},'' \href{http://dx.doi.org/10.3390/sym15040945}{{\em
  Symmetry} {\bfseries 15} (2023) 945},
  \href{http://arxiv.org/abs/2304.10789}{{\ttfamily arXiv:2304.10789
  [hep-th]}}.

\bibitem{Bertolini:2023sqa}
E.~Bertolini, N.~Maggiore, and G.~Palumbo, ``{Covariant fracton gauge theory
  with boundary},'' \href{http://dx.doi.org/10.1103/PhysRevD.108.025009}{{\em
  Phys. Rev. D} {\bfseries 108} no.~2, (2023) 025009},
  \href{http://arxiv.org/abs/2306.13883}{{\ttfamily arXiv:2306.13883
  [hep-th]}}.

\bibitem{Afxonidis:2023pdq}
E.~Afxonidis, A.~Caddeo, C.~Hoyos, and D.~Musso, ``{Fracton gravity from
  spacetime dipole symmetry},''
  \href{http://dx.doi.org/10.1103/PhysRevD.109.065013}{{\em Phys. Rev. D}
  {\bfseries 109} no.~6, (2024) 065013},
  \href{http://arxiv.org/abs/2311.01818}{{\ttfamily arXiv:2311.01818
  [hep-th]}}.

\bibitem{Rovere:2024nwc}
D.~Rovere, ``{Anomalies in Covariant Fracton Theories},''
  \href{http://arxiv.org/abs/2406.06686}{{\ttfamily arXiv:2406.06686
  [hep-th]}}.

\bibitem{Pretko:2016lgv}
M.~Pretko, ``{Generalized Electromagnetism of Subdimensional Particles: A Spin
  Liquid Story},'' \href{http://dx.doi.org/10.1103/PhysRevB.96.035119}{{\em
  Phys. Rev. B} {\bfseries 96} no.~3, (2017) 035119},
  \href{http://arxiv.org/abs/1606.08857}{{\ttfamily arXiv:1606.08857
  [cond-mat.str-el]}}.

\bibitem{Caddeo:2022ibe}
A.~Caddeo, C.~Hoyos, and D.~Musso, ``{Emergent dipole gauge fields and
  fractons},'' \href{http://dx.doi.org/10.1103/PhysRevD.106.L111903}{{\em Phys.
  Rev. D} {\bfseries 106} no.~11, (2022) L111903},
  \href{http://arxiv.org/abs/2206.12877}{{\ttfamily arXiv:2206.12877
  [cond-mat.str-el]}}.

\bibitem{Alvarez:2006uu}
E.~Alvarez, D.~Blas, J.~Garriga, and E.~Verdaguer, ``{Transverse Fierz-Pauli
  symmetry},'' \href{http://dx.doi.org/10.1016/j.nuclphysb.2006.08.003}{{\em
  Nucl. Phys. B} {\bfseries 756} (2006) 148--170},
  \href{http://arxiv.org/abs/hep-th/0606019}{{\ttfamily arXiv:hep-th/0606019}}.

\end{thebibliography}\endgroup

\end{document}